\documentclass[12pt,a4paper,aps]{article}
\usepackage{a4}
\usepackage{epsfig}
\usepackage[hmargin=2cm,vmargin=2cm,nohead]{geometry}
\begin{document}
\par
\title{The Future of Nuclear Energy: Facts and Fiction \\ 
Chapter IV: \\
Energy from Breeder Reactors and from Fusion?}

\author{
Michael Dittmar\thanks{e-mail:Michael.Dittmar@cern.ch},\\
Institute of Particle Physics,\\ 
ETH, 8093 Zurich, Switzerland}
\maketitle
\begin{abstract}
The accumulated knowledge and the prospects for commercial energy production from  
fission breeder and fusion reactors are analyzed in this report. 

The publicly available data from past 
experimental breeder reactors indicate that a large number of unsolved technological problems exist and that 
the amount of ``created"  fissile material, 
either from the U238 $\rightarrow$ Pu239 or from the Th232 $\rightarrow$ U233 cycle, is still far below the breeder requirements 
and optimistic theoretical expectations. 
Thus huge efforts, including many basic research questions with an uncertain outcome, are needed 
before a large commercial breeder prototype can be designed. 
Even if such efforts are undertaken by the technologically most advanced countries, 
it will take several decades before such a prototype can be constructed. 
We conclude therefore, that ideas about near-future commercial fission breeder reactors are nothing but wishful thinking. 

We further conclude that, no matter how far into the future we may look, nuclear fusion as an energy source is even less probable than 
large-scale breeder reactors, for the accumulated knowledge on this subject is already sufficient to say that commercial fusion power will never become a reality.
\end{abstract}


\newpage
\section{Introduction}

Over one hundred years ago, physicists began to understand that a huge amount of energy 
could be obtained from mastering nuclear fusion and fission energy. 
For example, the production of only 1 kg of helium from hydrogen ``liberates'' a thermal  
energy of about 200 million kWh. In the sun, this fusion reaction 
transforms about  600 million tons of hydrogen into helium every second, thus liberating   
$4 \times 10^{26}$ Joules per second.

The understanding of nuclear physics and its technological applications proceeded with a breathtaking speed.  
It took only seven years from the discovery of the neutron in 1931 to the observation of the  
neutron induced fission of uranium at the end of 1938. This was followed, on the 2nd of December 1942, 
by a sustained nuclear chain reaction with a power of 0.5 Watt (and up to 200 Watt at a later time) by E. Fermi and his team below the Chicago University football stadium~\cite{fermireactor}. 
The next steps in using nuclear energy were the explosions of the Hiroshima and Nagasaki fission bombs, on the 6th and 9th of August 1945, resulting in more than 100,000 deaths and the beginning of the nuclear arms race. Only  
a few years after the first fission bombs exploded, the USA and the Soviet Union had constructed 
hydrogen fusion bombs. These bombs were up to 1000 times more powerful than the Hiroshima fission bomb.
  
Also the peaceful application of nuclear fission energy happened very quickly: by 1954 
the thermal energy from a controlled fission chain reaction could be used 
to produce commercial electric energy \cite{firstelec}. During the next 30-40 years a large number of 
commercial nuclear power plants were constructed in most industrialized countries. 

The quick scientific and technical success in bringing 
this form of power into the production of commercial energy was impressive. 
Many nuclear pioneers expected that nuclear fission and fusion would provide their grandchildren with cheap, clean and essentially unlimited energy.  
In fact, these successes led most of us to a euphoric and blind belief in continuous scientific and technological progress. 
 
In contrast to such dreams,  nuclear fission energy nowadays is not cheap and 
even the most optimistic nuclear fusion believers do not expect the first commercial fusion reactor prototype 
until after 2050. 
One observes further that nuclear fission energy has been stagnating for about ten years  
and that its relative share in the worldwide electric energy production has decreased from about 18\% during the nineties to 
only 13.8\% \cite{ieanuclearfrac}.

Furthermore,  the average age of the existing nuclear power plants, the limitations of primary and secondary uranium 
resources as well as the problems related to nuclear proliferation and nuclear waste all lead to doubts  
about the prospects of the standard water moderated nuclear fission reactors. 
In fact it seems clear at this point that as fossil-fuel energy production 
declines, sufficient energy to ensure the survival of our highly industrialized civilization can not come 
from a rapid growth of nuclear fission energy of this sort.

The problem with the limited amount of uranium resources can theoretically be addressed with the mastering of the technology 
of nuclear fission breeder reactors.
It is claimed that this technology could increase the amount of fissile material from uranium 
by a factor of 60-100 and much more if the thorium breeder cycle can be realized \cite{breederfuel}.
It is believed that the breeder technology will enable us to bridge the time gap before 
nuclear fusion energy, which would become the ``final solution" to all energy worries, can be mastered \cite{bridgefusion}. 

In this final chapter IV of  ``The Future of Nuclear Energy" report, we discuss the experience with 
past and current breeder reactors in section 3. We analyze how the remaining problems will be addressed 
with the worldwide Generation IV breeder reactor program and with thorium based breeder reactors (section 4). 
The remaining obstacles towards a controlled and sustained nuclear fusion reaction chain are presented in section 5.     
In order to simplify the discussion, we start in section 2 
with some facts and basic physics principles of nuclear fission and fusion energy.

\section{Energy from nuclear fission and fusion, some facts and physics}

As we have discussed in detail in chapters I-III of this report \cite{chapter123}, 
the publicly available data on the long term worldwide natural uranium supply are in conflict 
already with a moderate annual 1\% growth rate for conventional water moderated reactors.

Consequently, believers in a bright future of nuclear energy should 
concentrate their efforts either (1) on the realization of the nuclear fuel breeder 
technology based on the uranium cycle, U238 to PU239, and the thorium cycle, 
TH232 to U233, or (2) on the mastering of the commercial nuclear fusion reaction.  

\noindent
In this section an overview of the existing and planned nuclear reactor types
and the experience with fast breeder reactors (FBR) is given (2.1).  
This is followed by a basic summary of the most important principles relevant for the use of 
nuclear fission and fusion energy (2.2 to 2.4).

\subsection{Facts on the existing and planned nuclear reactor types}

The worldwide nuclear fission reactors produced 2601 TWhe during the year 2008, or roughly 14\% of the world wide electric energy. 

For the year 2009 one finds that the commercial nuclear energy production will 
come from 436 nuclear fission reactors with a combined nominal electric power of 370.260 GWe\cite{pris}. 

\small{
\begin{table}[h]
\vspace{0.3cm}
\begin{center}
\begin{tabular}{|c|c|c|c|c|c|c|c|c|c|}
\hline
Reactor Type  & \multicolumn{3}{c|} {Terminated}     & \multicolumn{3}{c|} {Operating}             & \multicolumn{3}{c|} {under construction} \\ \hline
(IAEA/PRIS)       & \# & Power[GWe] &  \%         & \# & Power[GWe] & \% & \# & Power[GWe] &  \% \\
\hline
PWR                    & 34 & 15.6 & 43              &  264 & 243 & 66           &  43 & 39.8 & 84 \\ 
PHWR                 &   5 &   0.3 &  0.8            & 44    & 22.4 & 6.1         &    4 & 1.3 & 2.8 \\ 
BWR                    & 23 & 6.67 & 18             &  92   & 83.6 & 23          &    3 & 3.9 & 8.3 \\ 
other                    & 54 & 12.7 & 35             &  34  & 20.3  & 5.5         &    1  & 0.92  & 2   \\ \hline
FBR                     &   6  & 1.5  &   4.3          & 2      & 0.69& 0.2          &    2  & 1.2    & 2.6 \\ \hline
total                     & 122 & 36.7  & 100        & 436  & 370 & 100         & 53  & 47.2 & 100 \\ \hline
\end{tabular}\vspace{0.1cm}
\caption{The evolution of different reactor types and their corresponding 
electric power rating from the IAEA PRIS data base (October 2009) \cite{pris}. 
Another 5 reactors are listed in the ``Long Term Shutdown" category, four are PHWR's and one of them is the 
0.25 GWe Monju sodium cooled FBR reactor in Japan. 
}
\end{center}
\end{table}
}

The PRIS data base from the International Atomic Energy Administration (IAEA) shows that the dominant reactor type today and currently under 
construction are the water moderated fission reactors. The abbreviations PWR (PHWR) stand for 
pressurized (heavy) water reactors and BWR for boiling water reactors. 
As can be seen from Table 1, these 
reactors provide over 94\% of the nuclear fission power worldwide. 
The remaining 6\% of the nuclear fission power comes from graphite moderated and water or gas cooled 
older and smaller reactors.
It seems that the PWR type has won the competition for the existing reactors and for the next round 
by a large margin.  

One observes that only two FBR's are declared operational and one is since 1995 in a ``long term" shutdown phase.  
These FBR's contribute currently 0.2\% of the world nuclear power. 
This tiny contribution from FBR's today is even smaller than it used to be.
In the list of 122 terminated reactors one finds 6 FBR's with a combined power of 1.6 GWe, or 4.3\%.
In the list of 53 reactors (October 2009) currently under construction,
one finds only two relatively small FBR's. 

These numbers indicate not only that FBR's have a negligible role today 
and during the next 10 years, but also that their operation experience is far from being an
economical and technological success story. 
Some more details on the worldwide experience with various types of 
commercial FBR and "thorium fuel breeder" reactors and their operation are listed below: 

\begin{itemize}
\item The best operation experience comes from the BN-600
FBR reactor with a rated power of 0.56 GWe in {\bf Russia}.
This reactor has been operated commercially for 28 years and is scheduled 
to close in 2010 \cite{BN-600}. Its average energy availability is given as 73.79\%. In a specialized 
document from the IAEA fast reactor data base, \cite{iaeaFR}, one finds that this reactor would be better called 
a ``Fast Reactor", as it was designed to use more fuel than it could produce. 
A new BN-800 reactor with 0.8 GWe, is currently under construction in Russia and its scheduled start is now given as 2014. As its smaller ``brother", it is designed to consume Pu239 rather than to produce more fissile material. 
\item The other ``operating" FBR is the Phenix reactor in {\bf France}. Phenix originally 
started operation with a power of 0.233 GWe in 1974. Since 1997 it is rated with only 0.13 GWe 
and an energy availability factor of 60.23\% in 2008. According to the WNA (world nuclear association) data base it ceased 
power production in March 2009 and will continue a research program till October 2009 \cite{WNAphenix}. 
The larger Super Phenix reactor, with a power rating of 1.2 GWe,  
achieved a maximal energy availability of 32.6\% only. This very low performance, in comparison to PWR's, 
was achieved during the last operational year (1996) and after a short lifetime of 10 years.
\item  The Monju  reactor in {\bf Japan} was closed after a serious sodium leak in 1995. 
Since many years the reactor is scheduled to start the subsequent year. Perhaps this time it will really 
restart during the first few months of 2010 \cite{monjustatus}.  
\item The next generation FBR reactor is currently under construction in {\bf India}. According to the current plans, it will   
start producing electric energy during the year 2011 \cite{indiasbreeder}.
\item The KNK II reactor in {\bf Germany} is listed in the IAEA data base, \cite{iaeaFR}, with a tiny capacity of 0.017 GWe. 
During its operational lifetime, 1978 to 1991, it achieved an average energy availability factor of 23.65\%. 
A larger FBR, the SNR-300, with a rated power of 0.3 GWe was completed in 1985 but for various reasons 
never started. A large 1.5 GWe FBR, the SNR-2, never terminated even the design phase.  
\item A rather limited experience with a thorium admixture in the nuclear fuel in commercial prototype reactors exists. 
A WNA document mentions two THTR (Thorium High Temperature Reactor) \cite{WNAthorium}:  
One with 0.3 GWe in {\bf Germany},    
which operated commercially between 1986 and 1989; the second was the Fort St Vrain reactor 
with a power rating of 0.33 GWe in the {\bf USA}. It is listed 
as the only commercial thorium-fuelled nuclear plant, following closely the german design. 
It was operated between 1976-1989.

The WNA document mentions further that the experimental Shippingport reactor in the {\bf USA}, rated power of 0.060 GWe, has successfully demonstrated the concept of a Light Water Breeder Reactor (LWBR) using thorium. 
The Shippingport reactor began commercial electricity production in December 1957.
In 1965 the Atomic Energy Commission started designing the uranium-233/thorium core for the reactor. The reactor was operated as a LWBR between August 1977 and October 1982. 
\end{itemize}

Several countries have so far managed to 
construct GWe water moderated slow neutron reactors, mostly of the PWR type. These reactors were operated 
safely and efficiently for many years, using U235 fuel enriched to 3-4\%.    

In contrast, large breeder reactors, based on a large amount of initial fissile material and  
the transformation of U238 and Th232 for making new reactor fuel, 
have so far not even passed successfully a prototype phase. 

\subsection{Energy from nuclear fission and fusion, some basics}

Atoms consist of a nucleus, made of protons and neutrons, and electrons. 
The size and the chemical properties of atoms are defined by the number of electrons
surrounding the nucleus. 
The combined mass of the protons and neutrons, each 2000 times heavier than the 
electrons, defines roughly the mass of the atoms.
As the nucleus is 100000 times smaller than the atom
it follows that its mass density is huge in comparison with the atom. 
The same chemical characteristics can be expected for atoms with a fixed number of protons 
and with different number of neutrons and the energy in chemical reactions is of the order of  
1 eV ($1.6 \times 10^{-19}$ Joule).  
As the nuclear properties of an atom depend on the 
number of neutrons, the name isotope has been introduced to 
separate the chemically identical atoms according to their number of neutrons. 

Without going into details, it is known today that the 
energy source of the sun and other stars is nuclear fusion. 
This fusion starts from the large number of hydrogen atoms present in the sun.  
The fusion reaction in stars is possible because of  
the enormous gravitational pressure which overcomes the electric repulsive force 
between positively charged protons. Fusion is the
source of all heavier elements which were formed in super-novae explosions of super large early stars
and shortly after the big bang.  
For our subsequent discussions on nuclear fusion it is important to note that a 
relatively low fusion power density, about 0.3 Watt/m$^{3}$,
is found in the sun \cite{fusioninthesun}. In comparison, the power density envisaged for a hypothetical fusion reactor must be  
at least one million times larger.
 
The nucleus is bound by the very strong nuclear force, which acts against the repulsive 
electrostatic force of the protons. Measurements have shown that the mass of the 
various atoms is almost 1\% smaller than the mass of the individual protons and neutrons combined.
Following Einstein's famous $E=mc^{2}$ formula,  this mass defect corresponds to a huge amount of 
energy, about 8 MeV (8 million eV) per nucleon. 
This energy is liberated when one manages to fusion the different nucleons together. 
Starting from the different hydrogen isotopes, e.g. one proton, deuterium (one proton plus one neutron) 
and tritium (one proton plus two neutrons) a binding energy of up to a few MeV is found.  
Further fusion of these hydrogen isotopes into the helium nucleus liberates another roughly 20 MeV. 

Neutrons and protons in heavy atoms like uranium
are less strongly bound than in lighter atoms like iron and 
energy can be released in the fission of such heavy atoms. 
For example 1 MeV per nucleon, or 200 MeV in total, will be liberated in the 
fission process of U233, U235 and U238, each containing 92 protons and 141, 143 and 146 neutrons respectively. 
The energy liberated per fission reaction is at least 100 million times larger than in a chemical reaction.

It is therefore no wonder that this has created an enormous interest in subatomic physics and its 
application for ultimate weapons or for the commercial use of energy.    

\subsubsection{Civilian and military use of nuclear energy, some remarks} 

The focus of this report is the commercial use of nuclear energy.  
As the evolution of nuclear energy has always been strongly coupled with the military 
sector, we feel that a few remarks about the dangers of nuclear weapons and the ambiguity of 
the commercial use of nuclear energy are needed. 
First of all, governments wishing to have nuclear weapons 
were not faced with unsolvable problems related to the development of fission bombs based on Pu239 and U235.
This is especially true if nuclear physics and engineering know how 
had been built up under the umbrella of peaceful and commercial use of nuclear fission energy. 

Furthermore, it is interesting to notice that advocates of nuclear fission energy 
like to explain why the dangers from nuclear weapons are far less important than 
believed. This is usually followed by the statement that their praised
future nuclear energy technology will avoid proliferation 
problems\footnote{A similar contradiction in the argumentation is found 
with respect to safety and radiation issues. The existing nuclear power  
plants are claimed to be very safe and problems are small compared to many 
other dangers of modern life. However, when the favorite future nuclear 
energy system is introduced, it is always pointed out that the remaining risks will be further reduced by a large factor.}.

For example it is often argued that U233 produced in a future Th232 breeding cycle
will be useless for nuclear weapons. This argument is certainly flawed as countries who want to have 
the nuclear weapon capacity will most likely choose the simpler way to make a bomb using Pu239 or 
U235. Furthermore, those who know how to breed and separate hundreds of kg's of  
U233 can easily replace Th232 with U238 and produce a few tenth of kg of Pu239, sufficient to construct 
a few nuclear bombs. 

Those not yet convinced about the ambiguity between the peaceful and military application of the 
nuclear energy technology, should rethink their positions with respect to the 
Nuclear Proliferation Treaty, the NPT, and to the 
``evil" government of Iran. 

A careful reading of the treaty, \cite{NPT}, reveals that Iran, at least so far, is in agreement  
with the NPT obligations. However one finds 
that NPT member countries should not exchange nuclear knowledge 
with nuclear weapon countries outside the treaty.
It is also worth remembering that the official nuclear weapon states, Russia, USA, UK, France and China, 
have declared in the treaty their intention to eliminate nuclear weapons 
as quickly as possible. Almost forty years after these countries signed the NPT, they  
still have more than 20000 nuclear warheads.

The nuclear arms race at the end of the second world war and during the subsequent cold war is
well documented in many reports, books and movies and we refer to the 
extensive literature largely available now on the internet.
Especially for those who are not yet convinced about the dangers of nuclear weapons, we would like to recommend
the short youtube video on the largest explosion ever, the 60 Megaton hydrogen bomb in Siberia 
in 1961, \cite{zarbomb}, and to Stanley Kubrics masterpiece movie ``Dr. Strangelove, or how 
I learned to stop worrying and love the bomb" 
from 1964 \cite{strangelove}. This film, even though almost 50 years old, presents many still relevant ideas  
related to the 20000 remaining nuclear warheads.  

\subsection{Liberating the energy from nuclear fission and fusion}

As we have seen in the previous section, a large amount of energy per reaction can be liberated 
from the fusion of light elements and from the fission of heavy elements like uranium. 
However at least two additional conditions must be fulfilled before such a process 
can be considered for energy production.
\begin{itemize}
\item In order to obtain a useful amount of energy from nuclear reactions a continuous 
and controllable fission or fusion must be achieved for a large number of atoms. 
For example $10^{20}$ U235 atoms, (0.05 gr)\footnote{This amount of U235 is found in 6 gr 
of natural uranium.}, need to be split every second in a 1 GWe nuclear fission reactor.   
\item Enough raw material must be continuously available to sustain this chain reaction.   
\end{itemize}

Only three relevant isotopes fulfill these conditions for the nuclear fission process.
These are the two uranium isotopes U235 and U233 and the plutonium isotope Pu239. 
The energy liberated in the fission process is carried dominantly (about 80\%) by the two daughter atoms. 
This energy is relatively easily transferred to a liquid or gas and the heat can be used to 
operate a generator.

The chain reaction is possible as each neutron induced fission reaction produces 
on average between 2-3 neutrons. As one neutron is needed to initiate another 
fission reaction, 1-2 excess neutrons minus some inevitable losses are in principle available to  
increase the reactor power or perhaps to start a nuclear fuel breeding process. 
The introduction of neutron absorbers allows to control the reactivity of the nuclear reaction and thus  
to increase or decrease the reactor power. 

As we have seen in section 2.1, most of the large scale nuclear power plants of today comes from the PWR's (pressurized water reactors). They use dominantly U235 as primary reactor fuel. In these reactors the prompt fission neutrons, with kinetic energies of 1 MeV, are slowed down (moderated) by elastic collisions with the hydrogen nuclei in the water molecules to subeV kinetic energies. The nuclear fission probability with such slow neutrons is increased by a factor of up to several hundred. As a consequence 
a large reactor can be efficiently operated and controlled with a relatively low initial enrichment of U235 and
large scale power production with moderated neutrons has been mastered by many countries. 
The combined running experience of such large scale reactors, currently more than 13000 years, has resulted in stable 
electric energy production combined with 
small or negligible risks during the 
``regular operation"\footnote{The possibility to operate all reactors and for a long time
under regular safe conditions is questioned by many.}
up to an electric power output of more than 1 GWe. 

In comparison the neutron escape rate in smaller reactors 
and in unmoderated fast reactors is much higher. Therefore a 
chain reaction in FBR's with comparable reactor power is more difficult to control and 
a larger amount of initial fissile material with a higher density is needed.
One consequence is that the required technology to make such highly enriched nuclear fuel 
will always be faced with the problem of its dual use for bomb making.
 
The use of the excess neutrons for the transformation of the U238 and Th232 isotopes 
into fissile Pu239 and U233 looks very interesting as the
amount of fissile material could be increased theoretically by a factor of more than one hundred. 
The breeding reactions considered would use the excess neutrons according the two reactions: 

\begin{center}
$ U^{238}_{92}+n \rightarrow U^{239}_{92} (\beta^{-}$ decay) $\rightarrow Np^{239}_{93} (\beta^{-}$ decay) $ \rightarrow Pu^{239}_{94} $  and \\
\end{center}

\begin{center}
$ Th^{232}_{90}+n \rightarrow Th^{233}_{90} (\beta^{-}$ decay) $\rightarrow Pa^{233}_{91} (\beta^{-}$ decay)$ \rightarrow U^{233}_{92} $  \\
\end{center}
 
Some advantages and disadvantages for the U238 $\rightarrow$ Pu239 and 
the Th232 $\rightarrow$ U233 breeding cycles and some practical problems 
are listed in Table 2. Some of these problems and their proposed solutions will be discussed in detail in section 3 and 4 of this report. 
So far only little or no experience exists with large scale GWe breeder prototypes. 

\begin{table}[h]
\small
{
\begin{tabular}{|c|c|c|}
\hline
Problems and Advantages            &  U238 $\rightarrow$ Pu239  breeding &  Th232 $\rightarrow$ U233  breeding \\
\hline
average concentration (earth crust) &  2-3 ppm & 10 ppm \\ 
raw material availability  today      &  up to 2 million tons U238                   &  a few 1000-10000 tons(?)   \\
existing mining                               & about 40000 tons/year                      &  about 1000 tons/year(?) \\
\hline
computer based simulations           & no major problems                          & no major problems \\
max theoretical breeding gain                & (if initiated with PU239) 0.7      & 0.45 \\
required neutron spectrum             & fast (prompt MeV neutrons)             & fast to slow \\
half life of intermediate state    & NP239 (2.3 days half life)                 & Pa233 (27.4 days)  \\
intermediate neutron absorbers     & small  (?)                                           & large (Pa233)         \\
\hline
prototype experience                       & small scale to large scale                  & (one) small scale  \\
large scale operational experience   & (one) ``limited"                              & none  \\
breeding gain (reactor conditions)    & unclear (not 100\% public)                & 0.013 (after 5 years) \\
initial fission start up                          & U235 or Pu239         &  U233, U235 or Pu239 \\ 
fissile material fraction                       & $\geq$ 20\%       &  $\geq$ 20\%   \\
\hline
reactor cost relative to PWR              & huge                                                 &  comparable(?) \\ 
reactor lifetime relative to PWR         & small (so far)                                     &  comparable(?) \\ 

\hline
\end{tabular}\vspace{0.3cm} 
}
\caption{A qualitative comparison of the fissile breeding cycles with U238 and Th232.
The breeding gain is defined as the ratio of $\frac{C - D}{F} $ where C, D and F are the  
number of fissile atoms created, destroyed and fissioned.
In order to be called a breeder, more fissile material must be made than fissioned, 
and the breeding gain must be larger than zero.
The ``(?)" indicates guesstimate as good information has so far not been found by the author.
} 
\end{table}

We now turn to the fusion process.  Nuclear fusion can happen once the short range nuclear force
between nucleons becomes larger than the electrostatic repulsive force between two positively charged nuclei. 
This can happen if the protons involved either have large kinetic energies or if the 
protons are compressed by super large gravitational fields as observed in stars.  
Very high kinetic energies correspond to nucleus temperatures of many ten to hundred million degrees.
Such high kinetic energies can be obtained for example in accelerators but only for small numbers.
Larger amounts of fusion reactions can be obtained in special magnetic field arrangements.

It follows from first principles that the sometimes discussed ``cold fusion" reaction is 
in contradiction with our well established knowledge of subatomic physics. As the repulsive force 
increases with the number of protons involved, 
the conditions to achieve fusion with atoms heavier than hydrogen and its isotopes become more and more difficult.
It follows that fusion reactions based for example 
on the ``proton-boron" reaction and many others are only possible using accelerators. Ideas to use 
accelerators for continuous fusion reactions with commercially interesting GW power prove to be wishful 
thinking once the required amount of $10^{21}$ fusion reactions per second is considered. The very low efficiency for  
transforming electric energy into kinetic energy of proton beams poses another fundamental problem for such exotic ideas.

The probability for a fusion reaction depends on the product of the plasma temperature and 
the fusion reaction cross section. The deuterium-tritium fusion is a factor of 100 to 1000 easier to achieve than the next two fusion 
reactions of deuterium and $He^{3}_{2}$ and deuterium-deuterium respectively.  
As it is already extremely difficult to achieve even the lowest interesting plasma temperatures 
on the required large scale, it follows that the only possible fusion reaction under reactor conditions 
is the deuterium-tritium fusion into helium ($He^{4}_{2}$).

An additional advantage of this reaction is the fact that the produced additional neutron 
carries 14 MeV of the liberated energy of almost 18 MeV per fusion reaction out of the plasma zone.
Thus, in theory, it can be imagined that the 4 MeV carried by the helium nucleus are used to keep the 
plasma temperature high enough and that the neutron energy is transferred somehow to 
another cooling medium. This medium is imagined to transfer the heat to a generator.

Unfortunately tritium is unstable, its half life is only 12.3 years, and it does not exist in sizable amounts on our planet. 
It must therefore be produced in a breeding process. 
A possible chain reaction could follow this scheme: 

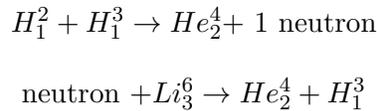
\begin{center}
$ H^{2}_{1} + H^{3}_{1}  \rightarrow He^{4}_{2} + $ 1 neutron \\
\end{center}

\begin{center}
neutron $+ Li^{6}_{3} \rightarrow He^{4}_{2} + H^{3}_{1}$ 
\end{center}

In comparison to the breeding and energy extraction in fission reactions, at least 
three additional fundamental problems can be identified for the fusion process. 
\begin{itemize}
\item A sustained super high temperature, at least 10 million degrees, is required 
in order to have fusion reactions happening at an interesting rate. Such high temperatures can be achieved in 
some special magnetic field arrangements or  
in a tiny volume with very intense laser or particle beams. Unfortunately  
no material is known which can survive the intense neutron flux under sustained reactor conditions 
and the sometimes occurring plasma eruptions.   
\item 
It is difficult to transfer the energy from the 14 MeV neutron  
to a gas or a liquid without neutron losses. 
\item The considered breeding reaction requires essentially that 100\% of the produced neutrons must be used 
to make tritium. As this is even theoretically impossible 
some additional nuclear reactions are proposed where heavier nucleons act as neutron multipliers.
However so far, even the most optimistic and idealized theoretical calculations have failed 
to produce neutrons in sufficient numbers.   
\end{itemize}

In short, the accumulated knowledge today indicates that the proposed fusion reaction is
unsustainable and can not lead to a sustainable power production. This statement will be strengthened   
with more details in section 5. 

\subsection{Dangers related to radioactive material}

We will conclude this section with some issues related to radioactive elements 
produced and liberated in the use of nuclear energy and the related 
dangers from ionizing radiation.
First of all, there are three types of radioactive decays, producing 
$\alpha$, $\beta$ and $\gamma$ radiation.
In addition, cosmic rays and various particles produced in high energy physics experiments 
should also be considered as a potential radiation hazard. 

The damage to cells is related to the ionizing potential or the energy deposit per volume originating from a source. 
The hazard is usually split into high and low radiation dose effects.
Very high radiation dose and the corresponding energy deposit result in fast cell death. 
If large and concentrated enough, the result can be the destruction of vital organs and death. 
It is important to know that the careless use of radiation during the early days of nuclear physics and its applications, 
have resulted in relatively high cancer rates among the participating scientists and engineers \cite{cancerradiation}.

The more tricky and less understood damage comes from small dose and long term effects to the 
cell DNA. While some self repair mechanism to broken DNA exists, it is also known  
that a single unlucky hit by a cosmic ray can transform the normal DNA into a cancer developing DNA, 
resulting many years later in the death of the host. It follows that 
the importance of small radiation doses for the development of a particular cancer types and in 
comparison to many other causes like smoking and asbestos is  
difficult to quantify. As a result, the associated cancer risks from small radiation doses 
will continue to fuel the emotional debate about nuclear energy for a long time. 

Despite these uncertainties, 
today the precautionary principle is used in many countries and very strict rules for people working in a radiation 
environment are applied. These rules are often summarized under the name 
ALARA (as low as reasonable achievable). The goal to reduce any radiation exposure to essentially negligible 
levels is one of the most important occupations of a radiation safety group. 
As a result of these efforts, assuming that expensive protection measures are taken, 
the health risks from radioactive contamination and under ``normal operation"  conditions 
are often much smaller than risks associated with working hazards in many other industrial domains. 
However, time pressure and profit optimization will always be 
in contradiction with the ever strengthened safety regulations.  

It is also known that it is essentially impossible to guarantee  the 
``normal operation" of the nuclear industry with its accumulating waste over periods of 
hundreds and sometimes even thousands of years. 
A solution to these problems is, as with similar long term problems 
of our industrial growth based societies,  left for future generations. 

\section{Experience with real breeder reactors} 

Breeder reactors are based on the beautiful idea 
that only one neutron, out of the 2.5 neutrons on average from the fission of U235 and U233 
(and 2.9 neutrons from Pu239), is required to keep the chain reaction going. 
It can thus be imagined, even if some neutron losses are allowed, 
that the additional neutrons can be used to make more nuclear fuel   
from U238 or Th232 than fissioned. Accordingly a reactor is defined 
as a breeder reactor if more fissile material is produced than consumed. 

The number of free neutrons per fission reaction
is $\eta = (\sigma_{f}/\sigma_{a}) \times v$, where $\sigma_{f}$ is the neutron induced fission cross section 
and $\sigma_{a}$ the neutron absorption (the sum of the neutron capture and fission) cross section and 
$\nu$ is the average number of prompt fission neutrons \cite{neutronphysics}. 
The fission to capture ratio and thus $\eta$ depend on the neutron energy and the different possible isotopes.  
As one neutron is required to sustain the chain reaction,  
breeding is only possible if $\eta$ is larger than 2. This condition is found 
for Pu239, U235 and U233 fission where 
$\eta$ for prompt  fast fission neutrons is  
2.7, 2.3 and 2.45 respectively. 
For thermal (moderated) neutrons U233 has the highest $\eta$ value of 2.3 followed by 
2.11 for Pu239 and 2.07 for U235.

Some Pu239 fuel production happens also in standard PWR reactors.
Depending on the reactor and fuel design characteristics as well as 
the amount of remaining fissile fuel in the reactor,  up to 30\% and more 
of the produced energy comes from the secondary Pu239 fission.

Two theoretical breeder options exist:

\begin{itemize} 
\item The use of thermal neutrons and the Th232 as input breeding material. 
\item The use of fast prompt neutrons dominantly from Pu239 fission, thus the name fast reactor, 
with U238 as the breeding material. 
\end{itemize}
 
The use of the Th232$\rightarrow$U233 cycle seems, at least on the first view, more attractive. 
The reaction can happen in the high fission cross section domain 
using moderated neutrons. The fission process with moderated neutrons 
is well understood, relatively easy to control and 
already in use with the standard nuclear water moderated reactors.
It seems that in principle one only needs sufficient amounts of U233 mixed with Th232 in order to start 
such a reactor and keep on going. Some of the remaining technical obstacles will be discussed in 
section 4.4. 

For the U238$\rightarrow$Pu239 breeder cycle one has to operate the fission 
process, either starting with U235 or Pu239, in the low fission cross section 
domain. As a consequence such reactors have to be operated with 
highly enriched U235 (HEU) or Pu239 fuel. Thus, one is not only confronted 
with special safety conditions for a large amount of "bomb" making material 
but also with a huge amount of fissile material which could 
under certain conditions reach the critical mass 
resulting in an uncontrolled chain reaction followed by a nuclear meltdown. 
Furthermore, the cooling of the active 
reactor zone has to be done with a low neutron absorption cross section and a high thermal 
conducting material like liquid sodium. Unfortunately, sodium is chemically very 
active and can easily burn in contact with oxygen.      

\subsection{The Shippingport LWBR thorium reactor}

The experience with the thorium breeder cycle comes mainly from research at 
the US Shippingport reactor, rated with a net power of 0.060 GWe. This reactor operated 
during the '60s, '70s and '80s. In 1965 the Atomic Energy Commission started designing the uranium-233/thorium core for the reactor. 
The reactor was operated as a LWBR between August 1977 and October 1982. 

According to the documentation, the reactor was started with 
a highly enriched 98\% U233 inventory of 501 kg and a total of 42260 kg of Th232 \cite{thoriumbreeder}.
No details are given about the origin of the 501 kg of U233.  However, one can  
assume that it came from a standard U235 fission reactor, where excess 
neutrons can be used to transform Th232 (or U238) blankets 
into U233 (or Pu239). 

The reactor had a maximum thermal power of 0.2366 MW(therm) and 
was operated for 29047 effective full hours, or about 66\% of the time. 
After five years of operation, a very detailed analysis of the fuel elements was performed.
It was found that the total U233 inventory had increased to 507.5 kg, a   
factor of 1.013. While it is impressive that the reactor could be operated and fueled with 
Th232 over a period of 5 years, the U233 gain was only about 6 kg of fissile material.

Assuming that such a reactor could eventually produce the U233 starting fuel 
for other reactors, one has to wait for quite some time before the second 
package of initial reactor core has been produced. Some large technological 
breakthroughs are required before this chain can be called feasible on a large scale.

The documents do not say much about the contamination of 
the 507.5 kg of U233 with fission products and its usefulness for further studies after this five year experiment.
The fact that no subsequent reactor experiment has been performed might provide a partial answer 
to this question.

Furthermore, it is interesting to note that the initial concentration of fissile material in a
reactor with only 0.237 GW(therm) energy was very large. It can be estimated that this amount, placed in a standard PWR,  
could have produced at least 5 times more electric energy than it had during the actual experiment. 

In contrast to the experiments performed at the Shippingport reactor, 
where the initial core was already U233, a realistic 
Th232 reactor cycle must be started with an initial U235 or Pu239 core.
Consequently, the experience gained with 
the Shippingport reactor experiment can not be considered as 
a proof that the envisaged system can function. It follows that many more 
tests are needed before one can imagine constructing a functioning large scale prototype 
Th232 breeder reactor.  

\subsection{Experience with Fast Reactors}

For the purpose of this report, the future of nuclear energy,  
we are mainly interested in the situation with the most important aspect, the question of the 
fuel breeding option.  Unfortunately very little information is provided for the 
experimental breeding achievements and most reports 
present the theoretical design breeding ratios.
For example the breeding ratio for the 
FBR Phenix reactor in France is given in many textbooks as 1.14\cite{breederratio}.
This number corresponds however to the theoretical design  
and it seems that a detailed experimental analysis, like the one done for the Th232 to U233 
cycle and the Shippingport reactor, is either secret or has not been performed.

Despite the missing experimental data of achieved breeding gains, the 
IAEA document, \cite{breedercores},
about the FBR core characteristics provides useful information about the design of such reactors. 
In this document a large number of FBR reactors, separated into 
(1) experimental Fast Reactors, (2) Demonstration of Prototype Fast reactors 
and (3) Commercial Size Reactors, are presented. 

The breeding gain, defined as the ratio of $\frac{C - D}{F} $ where C, D and F are the  
number of fissile atoms created, destroyed and fissioned, and other 
characteristics of different fast reactors are summarized in Table 3.

\begin{table}[h]
\small{
\begin{tabular}{|c|c|c|c|c|c|c|}
\hline
FR name  &  \multicolumn{2}{c|}{nominal Power [GW]}       &  \multicolumn{2}{c|}{fissile material core}        & enrichment & design             \\
(operation years)    &  therm &   elec  &  U235 [kg] &  Pu239$^{**}$    [kg]    & inner core   & breeding gain  \\
\hline
\multicolumn{7}{|c|}{Experimental Fast Reactors}  \\
\hline

Joyo                         & 0.14  &       -                      &  110    &   160                          & 30\%         & 0.03 \\
Fermi                       & 0.20    &    0.061               &   484    &      0                          & 25.6\%       & 0.16 \\
\hline 
\multicolumn{7}{|c|}{Demonstration or Prototype Fast Reactors} \\
\hline

Phenix                       & 0.563   &     0.255               &   35   &    931                         & 18.0\%       & 0.16 \\
SNR-300                       & 0.762 &       0.327               &  57  &  1536                       & 25.0\%       & 0.10 \\
PFBR$^{*}$                      & 1.250 &       0.500               &   17.3  &    1978                        & 20.7\%       & 0.05 \\
Monju                       & 0.714    &    0.280               &  13.5   &   1400                          & 16.0\%       & 0.2 \\
BN-600                       & 1.470    &    0.600               &  2020  & 112                            & 17.0\%       & -0.15  \\ 
\hline 
\multicolumn{7}{|c|}{Commercial Size Fast Reactors} \\
\hline

Super Phenix                      & 2.990  &      1.242               &   142 & 5780                & 16.0\%       & 0.18 \\
BN-800$^{*}$                       & 2.100   &     0.870               &   30    &        2710                    & 19.5\%       & -0.02 \\
\hline 
\multicolumn{7}{|c|}{Standard Water Moderated Reactors} \\
\hline

standard PWR          & 3.      &       1.                     &  3500      &      0$^{***}$                   &  3-4\%       & -0.7 \\
\hline

\end{tabular}\vspace{0.3cm} 
}
\caption{Some design values for the three groups of Fast Reactors, 
experimental, demonstration or prototype and commercial size \cite{breedercores}.
Reactors marked with a ``$^{*}$" are currently under construction. 
The design numbers can be compared with the ones of existing large commercial 
1 GWe PWR reactors, assuming an average charge of 500 tons of natural equivalent, given in the last line.
The ``$^{**}$" and ``$^{***}$" stand for a mixture of different plutonium isotopes dominated 
by Pu239 and the amount within the initial core respectively.  
}
\end{table}

It is very unfortunate that experimental breeding gains are not given in the IAEA fast reactor data base. 
In absence of any detailed publication, one can assume that the required detailed and very expensive 
isotope analysis of the reactor fuel has not been performed or published. 
The theoretical hopes for fuel breeding are thus not backed up with 
hard experimental data. 
Nevertheless, already the theoretical breeding gains of the different FBR's are 
revealing. Ten out of the twelve small experimental reactors
were operated in a {\bf configuration not for breeding}. 
The other two experimental reactors, listed in Table 3, are the Joyo in Japan 
and the Fermi in the USA. The Joyo reactor was not designed for the production of 
electric energy.  The Fermi reactor operated for a few years and 
had a partial core meltdown in 1966. 
This reactor was the first and only effort in the USA to operate a larger scale breeder reactor and was terminated in 1972. 

Another twelve Demonstration or Prototype reactors are listed in the IAEA report. Among them are the
Monju reactor in Japan, the ``Russian/Soviet"  BN-600 
and the Phenix reactor in France. 

Only the BN-600 reactor is currently operational and is often considered 
as the prime example of a successful operating FBR reactor. 
However, the IAEA document reveals that this reactor was designed with a negative 
breeding gain of -0.15.  

In comparison, the Phenix and Monju reactors are 
presented with a theoretical breeding gain of 0.16 and 0.2. 
It is interesting to note that the potential better construction of the next generation 
PFBR reactor in India, currently expected to start in 2011, 
is given with a much smaller theoretical breeding gain of 0.05.

The third FBR group in the IAEA document describes commercial size reactors. 
Eleven out of the listed thirteen large FBR's projects have been abandoned before any construction plans have been presented, 
or ``exist" currently only in the design phase.
Only one reactor, the Super Phenix reactor in France, has produced some electric energy.
During its short operation time it was operated with a very low  efficiency and it can not be considered as a successful breeder prototype. 
A new  commercial size fast reactor is under construction in Russia. The BN-800 is currently scheduled to become operational during the year 2014. 
It is however quantified with a negative breeding gain of -0.02.   

A further confirmation that the BN-800 reactor is not a breeder comes from 
WNA document, \cite{WNA-BN-800}, where the reactor is described as:

{\it ``It has improved features including fuel flexibility - U+Pu nitride, MOX, or metal, 
and with breeding ratio up to 1.3. However, during the plutonium disposition campaign 
it will be operated with a breeding ratio of less than one. 
"}

A possible interpretation of this statement could be that 
{\bf plutonium stocks are already a delicate problem and that Russia wants to get rid of it.}

In summary, the IAEA data base for fast reactors does not present any evidence 
that a positive breeding gain has been obtained with past and present FBR reactors. 
On the contrary, the presented data indicate at best that a more efficient nuclear fuel use 
than in standard PWR reactors can be achieved during normal running conditions.  
However, once the short and inefficient 
running times of FBR's, in comparison with large scale PWR's,  are taken into account, even this better fuel use has not been demonstrated.  
In fact, the required initial fuel load in FBR's contains at least twice as much 
natural uranium equivalent and with a fissile material enrichment of roughly 5 times larger than 
that in a comparable PWR. A fair comparison of the fuel efficiency should include 
the efficiency to recycle fissile material from used nuclear fuel in both reactor types.  

Three more areas of concern for a future breeder program should be added:
\begin{itemize}
\item  
Fast Reactors are known for a worrying safety record. For example, it might be true 
that serious incidents, like the one that happened with the Chernobyl graphite moderated reactor, can not happen 
with modern PWR's. However, only very few nuclear experts would sign such a statement for sodium cooled FBR's.   
\item 
FBR's are known for their huge construction costs relative to PWR's and it might be   
tempting to compare some of the past FBR's   
to a monetary ``black hole". An equivalent of  3.5 billion Euros
has been invested in the construction of the SNR-300 in Germany. Because of safety concerns related to sodium leaks and other 
problems, this small FBR has never started operation.   
This amount of money corresponds to the 
price tag for a five times more powerful modern PWR reactor.
\item 
A third problem is related to the FBR requirements to have a large inventory of fissile material combined with a high purity.
The amount of fissile material listed in Table 3 should be compared to a few kg required for a Pu239 bomb. 
This problem makes even small experimental reactors highly sensitive to 
the proliferation problem.
\end{itemize}

\section{Future Breeder Reactors} 

As our short overview in section 2 has already demonstrated, neither sodium 
cooled FBR reactors based on U238$\rightarrow$ Pu239 nor the Th232$\rightarrow$ U233 
cycle are fashinable commercial reactor types. 

As a consequence of the observation that known uranium deposits are 
limited, scientists from many countries have joined forces and created during the year 2001 the 
Generation IV reactor forum \cite{GENIV}. 
 
Within their own words (quote):
 
{\it ``The Generation IV International Forum, or GIF, was chartered in July 2001 to lead the collaborative efforts of the world's leading nuclear technology nations to develop next generation nuclear energy systems to meet the world's future energy needs."}

The work of over 100 experts from ten countries, including Argentina, Brazil, Canada, France, Japan, Republic of Korea, South Africa, Switzerland, the United Kingdom and the United States, and from the International Atomic Energy Agency and the OECD Nuclear Energy Agency has resulted at the end of the year 2002 
into a roadmap document with the title: \\

\begin{center}
{\bf A Technology Roadmap for Generation IV Nuclear Energy Systems} \\
\end{center}

After the definition of the goals, identifying promising concepts, their 
evaluation and the estimation of the required R\&D efforts, 
six systems have been selected. 
The selection was based on their estimation that 
they (quote) \\

{\it ``feature increased safety, improved economics for electricity production and new products such as hydrogen for transportation applications, reduced nuclear wastes for disposal, and increased proliferation resistance."}

Within the content of this analysis we are mainly interested whether the acknowledged
U235 fuel shortages can be solved with future breeder reactors. 
Therefore we will only take a closer look at the 
three FBR's and the one which has the potential to become a Th232 based thermal breeder.
According to the WNA document from August 2009 \cite{WNA-GenIV}: 
  
{\it ``at least four of the systems have significant operating experience already in most respects of their design, which provides a good basis for further R\&D and is likely to mean that they can be in commercial operation well before 2030."} \\

It is remarkable that the same WNA document contradicts this statement a few lines later: \\

{\it ``However, it is significant that to address non-proliferation concerns, the fast neutron reactors are not conventional fast breeders, i.e. they do not have a blanket assembly where plutonium-239 is produced.  Instead, plutonium production takes place in the core, where burn-up is high and the proportion of plutonium isotopes other than Pu-239 remains high.  In addition, new reprocessing technologies will enable the fuel to be recycled without separating the plutonium."}

\subsection{Some details about Generation IV breeder reactors}

The Generation IV roadmap document from the year 2002 describes a detailed planning 
for what needs to be achieved during the next 10-20 years. 
Depending on the results one might be able to decide 
which of the different future reactor concepts can be used to construct real prototype  
FBR's.

The qualitative proposed research plans for the three FBR's and the Th232 reactor can be summarized  
as follows:

\begin{itemize}
\item The Gas-Cooled Fast Reactor System (GFR). This concept is based on a helium-cooled  
reactor with a small thermal power of roughly 0.5 GW only. A large number of major technological
gaps are mentioned in the {\it Roadmap} leading to a research program of about 20 years and a cost of 
940 million US dollars.     
\item The Lead-cooled Fast Reactor System (LFR) with a possible thermal power between 0.1 GW and 3.6 GW.
A relatively long list of "technology gaps" for the LFR is presented, including even some insufficient knowledge 
of neutron interaction cross sections. A 15-20 year R\&D program with a price tag of 
990 million US dollars is needed before any further statements about the realization of this concept   
can be made. 
\item The Sodium-Cooled Fast Reactor System (SFR) with a thermal power rating between 1-5 GW.
This concept is closely related to the doubtful success  
with past sodium cooled fast reactors in France, Japan, Germany, the UK, Russia 
and the United States.
It is said that this reactor must be capable of also using the thermal neutron spectrum, because the 
startup fuel for the fast reactor must come ultimately from spent thermal reactor fuel.  
The list of technology gaps includes the need to ensure a ``passive safe response design basis", the 
``capital cost reduction" and the ``proof that a reactor has the ability to accommodate bounding events".
A somewhat frightening conclusion of this statement might be that 
previous sodium prototype FBR's did not fulfill 
any of these basic reactor safety standards. 

It is also mentioned that this sodium cooled reactor is the most advanced FBR system.  The required R\&D program 
to investigate the remaining problems could be done over a period of less than 15 years and for 610 million US dollar.
\item The Molten Salt Reactor system (MSR) is imagined as 1 GWe reactor with a net thermal efficiency of 44-50\%. 
The design assumes the use either U238 or Th232 as a fertile fuel dissolved as fluorides in the molten salt
and that it can operate with thorium as a thermal breeder. The technology gaps mentioned contain 
a large number of items related to the chemistry of molten salts as well as the need for 
more accurate basic neutron cross sections for compositions of molten salt.
The time scale of the required R\&D program is 15-20 years with an associated price tag of 1000 million dollars.  
\end{itemize} 

The Generation IV Roadmap document can be summarized with the statement that the known 
technological gaps to construct even prototype breeder reactors were enormous at the time when the 
document was written. These unknowns are addressed with 
a detailed planning for the required research projects and the associated cost.
Only after these problems have been solved a design and construction of expensive prototype breeder reactors can start. \\

We are now at the end of the year 2009 and almost half the originally planned R\&D period is over. 
Essentially no progress results have been presented and the absence of large funding 
during the past 8 years gives little confidence that even the most basic questions for the 
entire Generation IV reactors program can be answered during the next few years.
Thus, it seems that the Generation IV roadmap is already totally outdated and unrealistic.  

This is confirmed by 
the latest statements at the Global 2009 conference in September 2009 by B. Bigot, the chairman 
of the French Atomic Energy Commission, which indicate that the plan to have the reactors ready by the 
year 2030 is now delayed to 2040 and onwards. According to the Website "supporters of nuclear energy"
Bigot said {\it "from 2040 onwards, France was planning to use Generation IV FBRs when renewing its fleet."} \cite{Bigot2009}. 

\subsection{The Global Nuclear Energy Partnership (GNEP)}
Another initiative, the Global Nuclear Energy Partnership (GNEP), \cite{GNEP}, was  
announced by President Bush in his 2006 State of the Union address in 2006. 
In September 2007, all major nuclear energy countries, besides Germany and a few others 
have signed the statement of principles. According to the U.S. Department of Energy
the goals of the initiative are (quote):

{\it 
``First, reduce AmericaÕs dependence on foreign sources of fossil fuels and encourage economic growth. Second, recycle nuclear fuel using new proliferation-resistant technologies to recover more energy and reduce waste. Third, encourage prosperity growth and clean development around the world. And fourth, utilize the latest technologies to reduce the risk of nuclear proliferation worldwide." \\
}

However, in June 2009 the U.S. Department of Energy announced that it is no longer pursuing domestic commercial reprocessing, and had largely halted the domestic GNEP program. Research would continue on proliferation-resistant fuel cycles and waste management. \\

According to the WNA press information, \cite{WNA-GNEP}, the status of this initiative is: \\

{\it 
``Although the future of GNEP looks uncertain, with its budget having been cut to zero, the DoE will continue to study proliferation-resistant fuel cycles and waste management strategies. " \\
}

It follows that the GNEP initiative will not result in the construction of future breeder reactors.

\subsection{Ideas about using thorium as a reactor fuel}

During the past years a large number of articles and books, websites and blogs 
propose the use of thorium as the breeder material for future nuclear reactors \cite{Th232}. 
The promoters advocate many interesting possibilities which indicate that  
the Th232 cycle might have lots of advantages compared to the U238 breeder cycles in FBR's.  

The main problem with these wonderful new insights into the use of nuclear fission energy 
seems to be that nobody from the nuclear energy establishment is 
interested.
 
As a result, little or no private and public research money is invested into the question 
of how to develop a thorium breeder reactor. Ignoring the possibility that past investigations 
into the thorium fuel cycle have revealed several important problems, one needs to speculate 
about other reasons.   

\begin{itemize}
\item That the established nuclear energy experts 
do not like to see competition from outsiders. 
\item 
That the nuclear fusion community has managed to dominate 
the entire nuclear energy research domain and that the research budgets are already allocated to 
the ITER plasma research project. 
\end{itemize}

If these two possibilities contain some truth, those in favor of developing a thorium breeder reactor 
should start taking a strong position against the 
current nuclear energy establishment. They should point out that (1) the current use of nuclear energy 
has no perspective because of the limited amount 
of available uranium resources, (2)  the Th232 breeder cycle is by orders of magnitude 
better than the ideas about U238 breeder cycles with FBR's and  
(3) nuclear fusion is at least 50-100 years away. 
Leaving these more political issues aside,  we would like to repeat some  
rational statements and the otherwise rarely 
mentioned problems about the use of the Th232 breeder cycle from the WNA information article \cite{WNA-Th232}: \\

{\large {\bf Developing a thorium-based fuel cycle}}.

{\it ``In one significant respect U-233 is better than uranium-235 and plutonium-239, because of its higher neutron yield per neutron absorbed. Given a start with some other fissile material (U-233, U-235 or Pu-239) as a driver, a breeding cycle similar to but more efficient than that with U-238 and plutonium (in normal, slow neutron reactors) can be set up. (The driver fuels provide all the neutrons initially, but are progressively supplemented by U-233 as it forms from the thorium.) However, there are also features of the neutron economy which counter this advantage. 
In particular the intermediate product protactinium-233 (Pa-233) is a neutron absorber which diminishes U-233 yield."} \\

The statement continues with: \\
{\it ``Despite the thorium fuel cycle having a number of attractive features, 
development has always run into difficulties."
The main attractive features are:
\begin{itemize}
\item 
The possibility of utilizing a very abundant resource which has hitherto been of so little interest that it has never been quantified properly.
\item 
The production of power with few long-lived transuranic elements in the waste.
\item
Reduced radioactive wastes generally.
\end{itemize}

The problems include:

\begin{itemize}
\item 
The high cost of fuel fabrication, due partly to the high radioactivity of U-233 chemically separated from the irradiated thorium fuel. 
\item
Separated U-233 is always contaminated with traces of U-232 (69 year half-life but whose daughter products such as thallium-208 are strong gamma emitters with very short half-lives). Although this confers proliferation resistance to the fuel cycle, it results in increased costs.
\item 
The similar problems in recycling thorium itself due to highly radioactive Th-228 (an alpha emitter with two-year half life) present.
\item 
Some concern over weapons proliferation risk of U-233 (if it could be separated on its own), although many designs such as the Radkowsky Thorium Reactor address this concern.
The technical problems (not yet satisfactorily solved) in reprocessing solid fuels. However, with some designs, in particular the molten salt reactor (MSR), these problems are likely to largely disappear.
\item 
Much development work is still required before the thorium fuel cycle can be commercialized, and the effort required seems unlikely while (or where) abundant uranium is available. In this respect, recent international moves to bring India into the ambit of international trade might result in the country ceasing to persist with the thorium cycle, as it now has ready access to traded uranium and conventional reactor designs.
\end{itemize}
}

The WNA article concludes with the following diplomatic statement: \\

{\it ``Nevertheless, the thorium fuel cycle, with its potential for breeding fuel without the need for fast neutron reactors, holds considerable potential in the long-term. It is a significant factor in the long-term sustainability of nuclear energy."} \\

A ``logic" interpretation of the WNA statement and the list of arguments about thorium and within the context of our review could be:

\begin{itemize}
\item The breeding of Pu239 with fast neutrons has huge problems and it would be great if another nuclear fuel could be found.
\item Thorium breeding has an interesting potential if the remaining large number of problems can be mastered in the long term, but right now we are far away 
from this. The contamination with the strong neutron absorber Pa233 and the large radioactivity from U232 and other elements 
are chief among the currently unsolved problems.  
\item The well known use of nuclear fission energy in PWR's is unsustainable. The problems related to long-lived transuranic elements, e.g. plutonium and heavier elements as well as nuclear waste in general are unsolved.  The concern with nuclear weapon proliferation can not be solved either. 
\end{itemize}

\section{Fusion Illusions}
This section is a short version of a detailed article by the author in the second edition of the 
``The Final Energy Crisis" \cite{finalenergycrisis}. 

After the second world war, many nuclear pioneers expected that nuclear fusion would provide their grandchildren with cheap, clean and essentially unlimited energy. 
\par

Generations of physicists and physics teachers have been taught at the university, and have gone on to teach others, a) that 
progress made in fusion research is impressive, b) that controlled fusion is probably only a few decades away 
and c) that - given sufficient public funding - no major obstacles stand between us and success in this field.

Here are some quotes from physics textbooks that reflect this sort of optimism: \\

{\it ``The goal seems to be visible now''} (Nuclear and Particle Physics; Frauenfelder and Henley 1974) \\

{\it ``It will most likely take until the year 2000 to bring 
a laboratory reactor to full commercial utilization''} (Energy, Resources and Policy; R. Dorf 1978) \\

{\it ``As the construction of a fusion reactor implies a large number 
of unsolved practical problems, one can not expect that fusion will become a usable energy resource during some decades!
Within a longer time scale however it seems possible!''}
(Physics, P. A. Tipler 1991) \\

Obviously this has not happened yet; the fusion optimists have become more modest 
saying  {``if everything goes well, the first commercial 
fusion reactor prototype might be ready in 50 years from now"}. 
\par   
Such a statements only hide the fact that today 
we have no idea how to solve the remaining problems.   
The uncritical media of today waxed enthusiastically about the 
recent decision by the ``world's leaders'' to provide the ten billion dollars needed to start the ITER fusion project \cite{ITERhome}. 
During the past few years one could read, for example\cite{BBCnews}:

\begin{itemize}
\item ``If succesful, ITER would provide mankind with an unlimited source of energy"
(Novosti, 15. Nov. 2005).
\item
``Officials project that 10\% to 20\% of the world energy could come from fusion 
by the end of the century"  (BBC news 24 May 2006).  
\item ``If succesful, it could provide a source of energy that is clean and limitless." and 
``ITER says within 30 years, the electricity could be available on the grid!" 
(BBC news 21. Nov. 2006).
\end{itemize}

The public, worried about global warming and oil price explosions, seems to  
welcome the tacit message that ``we - the fusion scientists, the engineers, and the politicians - 
do everything that needs to be done to bring fusion energy on line before fossil fuel supplies become an issue, 
and before global warming boils us all.'' 

In what follows we challenge the assumption that the ITER project has any relation 
to the energy problem and we quantify the arguments of fusion sceptics. 

We start our discussion with an overview of the remaining huge problems facing commercial fusion
and give a detailed description of why  
the imagined self-sufficient tritium breeding cycle can not work.  
In fact, as we are about to see, it seems that enough knowledge has been accumulated on this subject to safely 
conclude that whatever might justify 
the 10 billion dollar ITER project, it is not energy research.

\subsection{Remaining barriers to fusion energy} 

Producing electricity from controlled nuclear fusion would require 
overcoming at least four major obstacles. The removal of each obstacle would need 
major scientific breakthroughs before any reasonable expectation might be formed of building 
a commercial prototype fusion reactor. 
It should be alarming that at best only the problems concerning the plasma control, described   
in point one below, might be investigated within the scope of the ITER project.
Where and how the others might be dealt with is anyone's guess. 

These are the four barriers:  
\begin{enumerate}
\item Commercial energy production requires steady state fusion conditions for a deuterium-tritium plasma  
on a scale comparable to that of today's standard nuclear fission reactors with outputs of 1 GW (electric) 
and about 3 GW(thermal) power. 
The current ITER proposal foresees a thermal power of only 0.4 GW using a plasma volume of 840 m$^{3}$.
Originally it was planned to build ITER with a plasma volume of 2000 m$^{3}$
corresponding to a thermal fusion power of 1.5 GW, but the fusion community soon realized  
that the original ITER version would never receive the required funding. 
Thus a smaller, much less ambitious version of the ITER project was proposed and finally accepted in 2005.

The 1 GW(el) {\it fission} reactors of today function essentially in a  
steady state operation at nominal power and with an availability time over an entire year of 
roughly 90\%. The deuterium-tritium {\it fusion} experiments have so far 
achieved short pulses of fusion power of 15 MW(therm) for one second and 4 MW(therm) for 5 seconds
corresponding to a liberated thermal energy of 5 kWh \cite{JET}
The Q value - produced energy over input energy - for these pulses was 0.65 and 0.2 respectively.
 
If everything works according to the latest plans \cite{ITERnew}, it will be 2018 when the first plasma experiments can start with ITER. 
From there it will take us to 2026, at least another eight years,  before the  first tritium experiments are 
tried\footnote{The original plans from 2005 are now, even before any serious construction has started, already delayed 
by four years.}.
In other words it will take at best 20 years from the agreement by the world's richest countries to construct ITER 
before one can find out if the goals of ITER,  
a power output of 0.5 GW(therm) with a Q value of up to 10 and for 400 seconds, are realistic. 
Compare that to the original ITER proposal which was 1.5 GW(therm), with a Q value between 10-15 and for about 
10000 seconds. 
ITER proponents explain that the achievement of this goal would already be an enormous success.
But this goal, even if it can be achieved by 2026, pales in comparison with the requirements of 
steady state operation, year after year, with only a few minor controlled 
interruptions. 

Previous deuterium-tritium experiments used only minor quantities of tritium and yet lengthy 
interruptions between successive experiments were required because the radiation from the tritium 
decay was so excessively high. In earlier fusion experiments, such as JET, the energy liberated in the short pulses 
came from burning (fusing) 
about 3 micrograms ($3\times 10^{-6}$ grams) of tritium, starting from a total amount of 
20 gr of tritium. 
This number should be compared with the few {\it kilograms} of tritium required to perform the experiments foreseen during the entire ITER lifetime
and still greater quantities that would be required for a commercial fusion reactor.
A 400 sec fusion pulse with a power of 0.5 GW corresponds to the burning  
of 0.035 gr ($3.5\times 10^{-2}$ grams) of tritium. A very large number when compared to 3 micrograms,
but a tiny number when compared with the yearly burning of 55.6 kilograms of tritium in a commercial 1 GW(therm) fusion reactor.

The achieved {\it efficiency} of the tritium burning (i.e., the amount that is burned divided by the total amount 
required to achieve the fusion pulse) was roughly 1 part in a million in the JET experiment and 
is expected to be about the same in the ITER experiments, far below any acceptable value if one wants to burn 
55.6 kg of tritium per year. 

Moreover, in a steady state operation the deuterium-tritium plasma will be ``contaminated'' with the helium nucleus that is produced 
and some instabilities can be expected. Thus a plasma cleaning routine is needed that would not cause noticeable interruptions of 
production in a commercial fusion plant. 
ITER proponents know that even their self-defined goal (a 400 second long deuterium-tritium 
fusion operation within the relatively small volume of 840 m$^{3}$) presents a great challenge.
One might wonder what they think about the difficulties involved in reaching {\it steady state} operation   
for a {\it full scale} fusion power plant.    
\item 
The material that surrounds and contains thousands of cubic meters of plasma in a full-scale 
fusion reactor has to fulfill two requirements. {\it First}, it has to survive an extremely high neutron flux 
with energies of 14 MeV, and second, it has to do this not for a few minutes 
but for many years.
It has been estimated that in a full-scale fusion power plant the 
neutron flux will be at least 10-20 times larger than in today's state of the art nuclear fission power plants.
Since the neutron energy is also higher, it has been estimated that -with such a neutron flux - each atom in the solid 
surrounding the plasma will be displaced 475 times over a period of 5 years~\cite{neutronerosion}.
{\it Second}, to further complicate matters, the material in the so called first wall (FW) around the plasma 
will need to be {\it very thin}, in order to minimize inelastic neutron collisions resulting in the loss of neutrons 
(for more details see next section), yet at the same time {\it thick enough} 
so that it can resist both the normal and the accidental collisions from the 100-million-degree hot plasma and for years.

The ``erosion'' for carbon-like materials from the neutron bombardment 
has been estimated to be about 3 mm per ``burn'' year
and even for materials like tungsten it has been estimated to be about 0.1 mm per burn year~\cite{neutronerosion}.

In short, no material known today can even come close to meeting the requirements described above. 
Exactly how a material that meets these requirements could be designed and tested remains a mystery, because 
tests with such extreme neutron fluxes can not be performed either at ITER or at any other existing or planned facility.

\item The radioactive decay of even a few grams of tritium creates radiation dangerous to living organisms, such that 
those who work with it must take sophisticated protective measures.
Moreover, tritium is chemically identical to ordinary hydrogen and as such very active and  
difficult to contain. Since tritium is also a necessary ingredient in hydrogen fusion bombs, there is additional risk that it might be
stolen. So, handling even few kg of tritium foreseen for ITER is likely to create major headaches both 
for the radiation protection group and for those concerned with the proliferation of nuclear weapons.  

Both of these challenges are essentially ignored in the ITER proposal and 
the only thing the protection groups have to work with today are design studies based on computer simulations. 
This may not be of concern to the majority of ITER's promoters today, since they will be retiring before the tritium 
problem starts in something like 10 to 15 years from now~\cite{ITERtimeline}. But at some point it will become a greater challenge 
also for ITER and especially once one starts to work on a real fusion experiment with many tens of kilograms of tritium.
\item Problems related to tritium supply and self-sufficient tritium breeding will be discussed in detail
in section 5.2. But first it will be useful to describe qualitatively 
two problems that seem to require simultaneous miracles if they are to be solved.

\begin{itemize}
\item The neutrons produced in the fusion reaction will be emitted essentially isotropically 
in all directions around the fusion zone. These neutrons must somehow be convinced to 
escape without further interactions through the first wall surrounding the few 1000 m$^{3}$ plasma zone. 
Next, the neutrons have to interact with a ``neutron multiplier'' material like beryllium in such a manner that the neutron flux is 
increased without transferring too much energy to the remaining nucleons.
The neutrons then must transfer their energy without being absorbed
(e.g. by elastic scattering) to some kind of gas or liquid, like high pressure helium gas, within the lithium blanket. 
This heated gas has to be collected somehow from the gigantic blanket volume and must flow to the outside.
This heat can be used as in any existing power plant to power a generator turbine. 
This liquid should be as hot as possible, in order to achieve reasonable efficiency for 
electricity production. However, it is known that the lithium blanket temperature can not be too high; this limits the efficiency 
to values well below those from today's nuclear fission reactors, which also do not have a very high efficiency.   
   
Once the heat is extracted and the neutrons are slowed sufficiently, they must make the inelastic 
interaction with the Li$^{6}$ isotope, which makes about 7.5\% of the natural lithium.
The minimal thickness of the lithium blanket that surrounds the entire plasma zone has been estimated 
to be at least 1 meter. Unfortunately, lithium like hydrogen (tritium atoms are chemically identical to hydrogen)  
in its pure form is chemically highly reactive. If used in a chemical bound state 
with oxygen, for example, the oxygen itself could interact and absorb neutrons, something that must be avoided. 
In addition, lithium and the produced tritium will  
react chemically - which is certainly not included in any present computer modeling - and some tritium atoms will be blocked within the blanket.
Unfortunately, additional neutron and tritium losses 
can not be allowed as will be described in more detail in section 5.2.
\item Next, an efficient way has to be found to extract the tritium quickly, and without loss, from this lithium blanket before it 
decays. We are talking about a huge blanket here, one that surrounds 
the few 1000 m$^{3}$ plasma volume. Extracting and collecting the tritium from this huge lithium 
blanket will be very tricky indeed, since tritium penetrates thin walls relatively easily, and since accumulations of tritium 
are highly explosive. (An interesting description of some of these difficulties that have already been encountered in a small scale experiment 
can be found in reference~\cite{anderson}.)
   
Finally assuming we get that far, the extracted and collected tritium and deuterium, which both need to be extremely clean, need to be
transported, without losses, back to the reactor zone. 
\end{itemize}
\end{enumerate}

Each of the unsolved problems described above is, by itself, serious enough to raise doubts about the success of 
commercial fusion reactors. But the self-sufficient tritium breeding is especially problematic, as will be described 
in the next section. 

\subsection{The illusions of tritium self-sufficiency}
The fact is, a self-sustained tritium fusion chain appears to be not simply problematic but 
absolutely impossible. To see why, we will now look into some details based on what is already known 
about this problem.

A central quantity for any fission reactor is its criticality, namely that exactly one neutron, out of the two to three neutrons 
``liberated'' per fission reaction, will enable another nuclear fission reaction.
More than 99\% of the liberated fission energy is taken by the heavy fission products such as barium and krypton 
and this energy is relatively easily transferred to a cooling medium. 
The energy of the produced fission neutrons is about 1 MeV.   
In order to achieve the criticality condition, the surrounding material must have a very low neutron absorption cross section
and the neutrons must be slowed down to eV energies.
For a self-sustained chain reaction to happen, a large amount of U235, enriched to 3-5\%,
is usually required. Once the nominal power is obtained, the chain 
reaction can be regulated using materials with a very high neutron absorption cross section.
A much higher enrichment of 20\% is required for fast reactors without moderators and up to 90\% for bombs.

In contrast to fission reactions, only one 14 MeV neutron is liberated 
in the $\mathrm{D + T \rightarrow He + n}$ fusion reaction. This neutron energy has to be transferred to a medium 
using elastic collisions. Once this is done, the neutron is supposed to make an inelastic interaction with a lithium nucleus,
splitting it into tritium and helium.      

Starting with the above reaction one can calculate how much tritium burning is required 
for continuously operating commercial fusion reactor assuming a power production of 1 GW(thermal)\footnote{This is relatively 
small compared to standard 3 GW(thermal) fission reactors which achieve up to 95\% steady state operation.}.  
One finds that about 55.6 Kg of tritium needs to be burned per year with an average thermal power of 1 GW.

Today tritium is extracted from nuclear reactors at extraordinary cost - about 30 million US dollar per kg from Canadian 
heavy water reactors. These old heavy water reactors will probably stop operation around the year 2025 and  
it is expected that a total tritium inventory of 27 kg will have been accumulated by that year~\cite{Abdou2003}.
Once these reactors stop operating, this inventory will be depleted by more than 5\% per year
due to its radioactive decay alone -tritium has a half-life of 12.3 years.    
As a result, for the prototype ``PROTO'' fusion reactor, which fusion optimists imagine to start operation not before the year 2050, 
at best only 7 Kg of tritium might remain for the start (Normal fission reactors produce at most 2-3 Kg per 
year and the extraction costs have been estimated to be 200 million dollars per kg~\cite{Abdou2003}.). 
It is thus obvious that any future fusion reactor experiment 
beyond ITER must not only achieve tritium self-sufficiency, it must create more tritium than it uses if 
there are to be any further fusion projects. 
  
The particularly informative website of professor Abdou from UCLA, one of the world's leading experts 
on tritium breeding, gives some relevant numbers 
both about the basic requirements for tritium breeding and 
the state of the art today~\cite{abdouwww}.

But first things first: 
Understanding such ``expert'' discussions requires an acquaintance with some key terms:

\begin{itemize}
\item The ``required tritium breeding ratio'', rTBR, stands for the minimal number of tritium nuclei which must be 
produced per fusion reaction in order to keep the system going. It must be larger than one, because of 
tritium decay and other losses and because of the necessary inventory 
in the tritium processing system and the stockpile for outages and for the startup of other plants.
The rTBR value depends on many system and technology parameters. 
\item The ``achievable tritium breeding ratio'', aTBR, is the value obtained from 
complicated and extensive computer simulations - so-called 3-dimensional simulations - of the blanket with its lithium and other 
materials. The aTBR value depends on many parameters  
like the first wall material and the incomplete coverage of the breeding blanket.     
\item Other important variables are used to define quantitatively the value of the rTBR. These include: 
(1) the ``tritium doubling time'', the time in years required to double the original inventory;
(2) The ``fractional tritium burn up'' within the plasma, expected to be at best a few \%; (3) The ``reserve time'',
the tritium inventory required in days to restart the reactor 
after some system malfunctioning with a related tritium loss; and (4) The ratio between the calculated and the experimentally obtained TBR.
\end{itemize}

The handling of neutrons, tritium and lithium requires particular care, not only because of  
radiation, but also because tritium and lithium atoms are chemically very reactive elements. 
Consequently, real-world, large-scale experiments are difficult to perform and 
our understanding of tritium breeding is based almost entirely on 
complicated and extensive computer simulations, 
which can only be done in a few places around the world.   

Some of these results are described in a publication by Sawan and Abdou from December 2005~\cite{sawanabdou05}. 
The authors assume that a commercial fusion power reactor of 1.5 GW (burning about 83 kg of tritium per year)
would require a long-term inventory of 9 kg and they further assume that the required start-up tritium is available.  

They argue that according to their calculations, the absolute minimum rTBR is 1.15, assuming a doubling time
of more than 4 years, a fractional tritium burn-up larger than 5\%
and a reserve time of less than 5 days. Requiring a shorter doubling time of 1 year, their calculations indicate that
the rTBR should be around 1.5.
Other numbers can be read from their figures.  For example
one finds that if the fractional burn-up would be 1\%\footnote{The fractional tritium burn-up during the short MW pulses in JET was roughly 0.0001\%.}
the rTBR should be 1.4 for a 5 year doubling time and
even 2.6 for a 1 year doubling time.
  
The importance of short tritium doubling times can be understood easily using the following 
calculation. Assuming these numbers can be achieved and that 27 kg tritium (2025) minus the 9 kg long term inventory,  
would be available at start-up, then 18 kg could be burned in the first year. 
A doubling time of 4 years would thus mean that such a commercial 1.5 GW(thermal) reactor can operate 
at full power only 8 years after the start-up.

And if anything these rTBR estimates are far too optimistic since a number of potential losses related to the 
tritium extraction, collection and transport are not considered in today's simulations.
 
The details become even more troubling when we turn to the tritium breeding numbers 
that have been obtained with computer simulations.

After many years of detailed studies, current simulations show that the blanket designs of today have, at best, achieved TBR's of 1.15. 
Using this number, Sawan and Abdou conclude that theoretically a small window for tritium self-sufficiency still exists. 
This window requires (1) a fractional tritium burn up of more than 5\%, (2) a tritium reserve time of less than 5 days and 
(3) a doubling time of more than 4 years. But using these numbers, the authors believe it is difficult even to imagine
 a real operating power plant. In their words, ``for fusion to be a serious contender for energy production, shorter 
doubling times than 5 years are needed'', and the fact is, doubling times {\it much shorter than 5 years} appear to 
be required, which means TBR's much higher than 1.15 are necessary. To make matters worse, they also acknowledge that 
current systems of tritium handling need to be explored further. This probably means that the tritium extraction 
methods from nuclear fission reactors are nowhere near meeting the requirements.

Sawan and Abdou also summarize various effects which reduce the obtained aTBR numbers once
more realistic reactor designs are studied and structural materials, gaps, and first wall thickness are considered.
For example they find
that as the first wall, made of steel, is increased by 4 cm starting from a 0.4 cm wall, the aTBR drops by about 16\%. 
It would be interesting to compare these assumptions about the first wall with the ones used in previous plasma physics experiments like JET
and the one proposed for ITER. Unfortunately, we have so far not been able to obtain any corresponding detailed information. 
However, as it is expected that the first wall in a real fusion reactor will erode by up to a few mm per fusion year,
the required thin walls seem to be one additional impossible assumption made by the fusion proponents.

Other effects, as described in detail by 
Sawan and Abdou~\cite{sawanabdou05}, are known to reduce the aTBR even further. 
The most important ones come from the cooling material required to transport the heat away from the breeding zone, 
from the electric insulator material, from the incomplete angular coverage of the inner plasma zone with a volume 
of more than 1000 m$^{3}$ and from the plasma control requirements. 

This list of problems is already very long and shows that the belief in a self-sufficient 
tritium chain is completely unfounded.
However, on top of that, some still very idealized TBR experiments have been performed now.
These real experiments show, according to Sawan and Abdou~\cite{sawanabdou05}, 
that the measured TBR results are consistently about 15\% lower than the modeling predicts. 
They write in their publication:
``the large overestimate (of the aTBR) from the calculation is alarming and implies that an 
intense R\&D program is needed to validate and update .. our ability to accurately predict 
the achievable TBR.''

One might conclude that a correct interpretation could have been: \\

Today's experiments show consistently that no window for a self-sufficient tritium breeding currently exists 
and suggest that proposals that speak of future tritium breeding are based on nothing more than hopes,
fantasies, misunderstandings, or even intentional misrepresentations.

\subsection{Ending the dreams about controlled nuclear fusion}

As we have explained above, there is a long list of fundamental problems 
concerning controlled fusion. Each of them appears to be large enough to raise serious doubts 
about the viability of the chosen approach to a commercial fusion reactor and thus about the 10 billion dollar ITER project.

Those not familiar with the handling of high neutron fluxes or the 
possible chemical reactions of tritium and lithium atoms might suppose that these problems 
are well known within the fusion community and are being studied intensively.
But the truth is, none of these problems have been studied intensively and, at best, even with the ITER project, 
the only problems that might be studied relate to some of the plasma stability issues outlined in section 5.1.
{\it All of the other problem areas are essentially ignored in today's discussions among ``ITER experts''.} 
 
Confronted with the seemingly impossible 
tritium self-sufficiency problem that must be solved before a commercial fusion reactor is possible, 
the ``ITER experts'' change the subject and tell you that this is not a problem for their ITER project.
In their view it will not be until the next generation of experiments - experiments that will not begin 
for roughly another 30 years according to official plans - that issues related to tritium self-sufficiency 
will have to be dealt with. Perhaps they are also comfortable with the fact that neither the problems 
related to material aging due to the high neutron flux nor the problems related to tritium and lithium handling 
can be tested with ITER. Perhaps they expect miracles from the next generation of experiments.
   
However among those who are not part of ITER and those who do not expect miracles, it seems that times are 
changing. More and more scientists are coming to the conclusion 
that commercial fusion reactors can never become reality. Some are even receiving a little attention from the media 
as they argue louder and louder that the entire ITER project has nothing to do with energy research~\cite{lemondeetc}.

One scientist who should be receiving more attention than he is, is Professor Abdou.
In a presentation in 2003 that was prepared on behalf of the US fusion chamber technology community 
for the US Department of Energy (DOE) Office of Science on Fusion Chamber Technology he wrote
that ``Tritium supply and self-sufficiency are `Go-No Go' issue for fusion energy, [and are therefore] 
as critical NOW as demonstrating a burning plasma'' [capitalization in original].
He pointed out that ``There is NOT a single experiment yet in the fusion environment that shows that the DT fusion fuel 
cycle is viable. He said that ``Proceeding with ITER makes Chamber Research even more critical'' and he asked
{\it ``What should we do to communicate this message to those who influence 
fusion policy outside DOE?''}~\cite{abdoudoe}. In short, to go ahead with ITER without addressing these chamber technology 
issues makes no sense at all. 
In light of everything that has been said in this section, it seems clear that this is what should be done:

Tell the truth to the tax payers, the policy makers and to the media; 
tell them that, after 50 years of very costly fusion research conducted at various locations 
around the world, enough knowledge exists to state:

\begin{enumerate}
\item that today's achievements in all relevant areas are still many orders of magnitude away from the 
basic requirements of a fusion prototype reactor; 
\item that no material or structure is known which can withstand the extremely high neutron flux
expected under realistic deuterium-tritium fusion conditions; and 
\item that self-sufficient tritium breeding appears to be absolutely impossible to achieve 
under the conditions required to operate a commercial fusion reactor.
\end{enumerate}

It is late, but perhaps not too late, to acknowledge that the ITER project is at this point 
nothing more than an expensive experiment to investigate some fundamental aspects of plasma physics. 
Since this would in effect acknowledge that the current ITER funding process 
is based on faulty assumptions and that ITER should in all fairness be funded on equal terms 
with all other research projects, acknowledging these truths will not be easy. But it is the only honest thing to do.

It is also the only path that will allow us to transfer from ITER to other more promising research 
the enormous resources and the highly skilled talents that need now to be brought to bear on our increasingly urgent energy problems. 
In short, this is the only path that will allow us to stop ``throwing good money after bad'' and to start dealing with our emerging 
energy crisis in a realistic way.  

\section{Summary}   

In this final chapter IV of our analysis about the ``Future of Nuclear energy" we have presented 
the status and prospects for nuclear fuel breeder fission reactors and the situation with 
nuclear fusion. 

Despite the often repeated claims that the technology for fast reactors is well understood, 
one finds that no evidence exists to back up such claims. In fact 
their huge construction costs,  their poor safety records and their inefficient performance give little 
reason to believe that they will ever become commercially significant. 

Indeed, no evidence has been presented so far that the original goal 
of nuclear fuel breeding has been achieved. The design and running plans 
for the two FBR's, currently under construction in India and in Russia do not 
indicate that successful breeding can even in principle be achieved.

Nevertheless, assuming that huge and costly efforts are made during the next 20-30 years, a remote possibility 
of mastering of nuclear fission breeder reactors can still be imagined. However, it is unclear if 
(1) enough highly enriched uranium remains to start future commercial breeder reactors on a large scale in 30-40 years from now
and (2) if the people in rich societies will accept risky and costly research efforts during times of economic difficulties.  
In any case Fast Breeder reactors even with the most optimistic assumptions will come far too late to 
compensate for the coming energy declines following the peak of oil and gas. 

In contrast to remaining open questions with fission breeders, 
we find that the accumulated knowledge about nuclear fusion is already large enough to conclude 
that commercial fusion power is not only 50 years away but that it will always be 50 years away.

The current situation concerning the future of nuclear energy appears in many respects similar to the one 
described in a famous fairy tail, ~\cite{hcanderson}, but with a slightly modified ending: 

\large{
{\bf{``In the coming ``autumn and winter" of our industrial civilization brought on by the decline of fossil fuels, it seems clear that 
the clothes of the {\it Nuclear Fission Energy} emperor are far too thin to keep him and others warm 
and that the {\it Nuclear Fusion Emperor} has no clothes at all!''}}. }

 \newpage
   
\noindent
{\bf \large Acknowledgments} 

\normalsize
This report and especially chapter IV about the ``Future of Nuclear Energy: Facts and Fiction" is a result of 
many unanswered questions which the author asked over the 
past few years directly to scientists active within the fission and fusion research community. 
Essentially none was answered and essentially no help was provided to get in contact with 
the corresponding ``fission" and ``fusion'' experts. Thus, in some kind of ``hobby'' research, which included 
discussions with friends, colleagues and many believers in a never ending technological progress,  
the different pieces concerning the future of nuclear energy summarized in this report came together.
 
During early 2007, an attempt was made to discuss the fusion problems 
in an open and scientific way directly with scientists from the fusion community. After coming as far as fixing the date for a seminar, 
the author received an email saying that there was a misunderstanding and a dialog did not take place. 
A similar approach, to discuss the open issues about nuclear fission energy 
was tried during 2008.  Again, it came as far as a seminar invitation canceled when trying to fix a date. 

However, during the spring 2007 the author received an invitation to present the ``Status and prospects of nuclear energy" at the 6th ASPO meeting in Cork, 
Ireland in September 2007. In preparation for this presentation the author took the time to study the 2005 edition of the Red Book in detail. 
Many questions about the uranium resources numbers, presented in the Red Book, came up but the inconsistencies were not yet large 
enough to start doubting the data. This view changed however when the 2007 edition appeared together with an enthusiastic press declaration in June 2008. 
As it turned out from the comparison of the 2007 and 2005 editions, the reported uranium resource data were nothing else than a collection 
of proven and unproven geological data mixed with some politically correct wishful thinking 
about a sustainable and bright future for the peaceful use of nuclear energy. 
This is how this report with its first three chapters about the Red Book and the analysis of future nuclear energy technologies started to take shape. 

Even though the views expressed in this paper are from the author alone, 
I would like to thank several colleagues and friends who took the trouble to discuss the content of this report during the past few years
with me. They all helped me to bring it into its final form. I would like to thank especially D. Hatzifotiadou, W. Tamblyn and F. Spano 
for many valuable suggestions and the careful reading of the paper draft. 
I would also like to thank S. Newman, who had asked me during the spring of 2007, to prepare a chapter about ``Fusion Illusions" 
for the second Edition of the book ``The Final Energy Crisis". Her encouragement was essential to write the longer report about nuclear 
fusion energy. 

Finally, after several attempts to complete also the report about the Red Book and the status and prospects of nuclear fission energy, it was 
Prof. F. Cellier who suggested to split this report into several chapters and submit it to the Oil Drum for publication. 
I am very grateful to him about the many valuable discussions we had, for the encouragement 
to complete this report and for his editing work to transform the article into the style  
needed for the oil drum publication. I am also grateful to the staff of the oil drum for having created a place where 
such articles, often censored in other places, can be published and confronted directly to the comments of a large number of critical readers.

Thus, the author hopes, with the ideas expressed in the quote from Gustave Le Bon below, 
that this report will function like some kind of ``telescope'', helping others to observe that some objects are moving around the Jupiter. 

\begin{center}
{\bf 
``Science promised us truth, or at least a knowledge of such relations as our intelligence can seize:
it never promised us peace or happiness'' \\
Gustave Le Bon }
\end{center}


\begin{thebibliography}{99}
\bibitem{breederratio}{The breeding ratio of 1.14 for the Phenix FBR is given in many papers and textbooks, however 
according to the \cite{breedercores} document this value is the design value and not the result of an experimental analysis.}
\end{thebibliography}
\end{document}